\documentclass[fleqn,usenatbib]{mnras}

\usepackage{newtxtext,newtxmath}

\usepackage[T1]{fontenc}
\usepackage{ae,aecompl}

\usepackage{graphicx}	
\usepackage{amsmath}	
\usepackage{amssymb}	
\usepackage{gensymb}
\usepackage{multirow}

\title[Fragmentation modelling of the August 2019 impact on Jupiter]{Fragmentation modelling of the August 2019 impact on Jupiter}

\author[R Sankar et al.]{
Ramanakumar Sankar,$^{1}$\thanks{E-mail: rshankar2012@my.fit.edu}
Csaba Palotai,$^{1}$
Ricardo Hueso,$^{2}$
Marc Delcroix,$^{3}$ \newauthor
Ethan Chappel,$^{4}$
Agustin S\'anchez-Lavega$^{2}$
\\
$^{1}$Department of Aerospace, Physics and Space Sciences, Florida Institute of Technology, Melbourne, Florida, USA\\
$^{2}$Depto. F{\'i}sica Aplicada I, Escuela de Ingenier{\'i}a de Bilbao, UPV/EHU, Bilbao, Spain \\
$^{3}$Commission des Observations plan{\'e}taires, Soci{\'e}te Astronomique de France, Toulouse, France \\
$^{4}$ Department of Computer Science, University of Texas at San Antonio, San Antonio, Texas, USA
}

\date{Accepted XXX. Received YYY; in original form ZZZ}

\pubyear{2020}

\begin{document}
\label{firstpage}
\pagerange{\pageref{firstpage}--\pageref{lastpage}}
\maketitle

\begin{abstract}
On 7th August 2019, an impact flash lasting $\sim1$s was observed on Jupiter. The video of this event was analysed to obtain the lightcurve and determine the energy release and initial mass. We find that the impactor released a total energy of $96-151$ kilotons of TNT, corresponding to an initial mass between $190-260$ metric tonnes with a diameter between $4-10$m. We developed a fragmentation model to simulate the atmospheric breakup of the object and reproduce the lightcurve. We model three different materials: cometary, stony and metallic at speeds of $60$, $65
$ and $70$ km/s to determine the material makeup of the impacting object. The slower cases are best fit by a strong, metallic object while the faster cases require a weaker material.
\end{abstract}




\section{Introduction}
Among the planets in the Solar System, Jupiter has the most intense gravitational field and the largest effective cross section. Giant impacts caused by objects of sizes 2 to 0.5 km have been observed in July 1994 \citep[the famous series of impacts from the fragments of the comet Shoemaker-Levy 9;][]{Harrington2004} and July 2009 \citep{SanchezLavega2010} respectively. These large impacts leave large aerosol debris fields in the planet's atmosphere visible for weeks to years \citep{Hammel1995,Hammel2010,SanchezLavega1998,SanchezLavega2011}. Impacts by smaller objects (sizes around 10-m) occur far more often and the impacts of 5 objects have been observed as short flashes of light on Jupiter with a duration about 1 second and peak brightness comparable or smaller than the Galilean satellites. These impacts were discovered by amateur astronomers operating small telescopes and using fast video cameras \citep{Hueso2010,Hueso2013,Hueso2018a}. The first of these small impacts occurred in June 2010 and was simultaneously observed by two observers using color filters in red and blue light. \citet{Hueso2010} analyzed this event obtaining calibrated light curves from the 1s flash. The conclusion was that the flash event was caused by the impact of an object with a diameter in the range of 10 m, releasing an energy comparable to an object of about 30 m impacting Earth's atmosphere. Later events have been analyzed by \citet{Hueso2013} and \citet{Hueso2018a}. A review of these impacts is available in \citet{Apostolos2019}. \citet{Hueso2018a} present a comparative study of all previous flash impacts on Jupiter and suggest an impact rate of similar objects of about 10-65 impacts per year with only a fraction of them being potentially observable from Earth in a perfect continuous survey of the planet (4-25 impacts per year). 

The discovery of new small impacts on Jupiter requires numerous observations with a cadence high enough to detect short flashes of 1-2 seconds and at least a modest quality atmospheric seeing. For instance, if the impact rate of these objects is high and 25 impacts occur each year in the visible side of Jupiter, discovering a new impact would require about 100 hours of observation time for a probability of 30\% of discovering a new impact. Such observations at 30 frames per second would imply the search of a faint brief flash in 10 million frames. However, if the impact rate is in the lower range above, a probability of 30\% of discovering a new impact would require acquiring observations over 600 hours, which is clearly not possible for any existing observing facility. Only the ensemble of amateur observers can observe Jupiter for such a long period of time and new impacts in the planets will most likely continue to be found by amateur observers. However, the flashes are brief and faint and could be unnoticed by an observer after hours of observing the planet. 
In fact, as reported in \citet{Hueso2018a}, most observers of previous impacts did not see the flash initially and only found it later when visually reviewing their data in some cases many days later.

Since 2012 we have developed a software tool called DeTeCt that performs differential photometry of video observations of Jupiter and can identify flash impacts in the planet. The development of the software was partially funded by the Planetary Space Weather Services of the Europlanet-2020 Research Infrastructure and is maintained by M.D. in a dedicated website\footnote{ \url{http://www.astrosurf.com/planetessaf/doc/project_detect.php}}. DeTeCt is documented in \citet{Hueso2018b} and is now widely used by around 100 observers who run the software over more than 123,000 video observations equivalent to 3240 hours of observation. 

On August 7th 2019, one of the observers who run DeTeCt regularly used the software over several videos acquired that night. The software identified a new impact in the planet, the sixth flash impact on Jupiter. The impact was recorded with an excellent seeing and at a fast frame rate of 82 frames per second allowing to retrieve the best-quality light curve of flash impacts on Jupiter. Here we describe the initial observations, the analysis of this impact and models that can fit the characteristics of the light-curve of the video observation. In section~\ref{sec:obs} we present the observations and the light curve of this event. In section~\ref{sec:massest} we calculate the energy released by the impact and a mass estimate of the impactor. In section~\ref{sec:fragmodel} we present a fragmentation model that is used to fit the light curve. We present and discuss the results of that fit in section~\ref{sec:results}.

\section{Observations}

\begin{figure}
    \centering
    \includegraphics[width=\columnwidth]{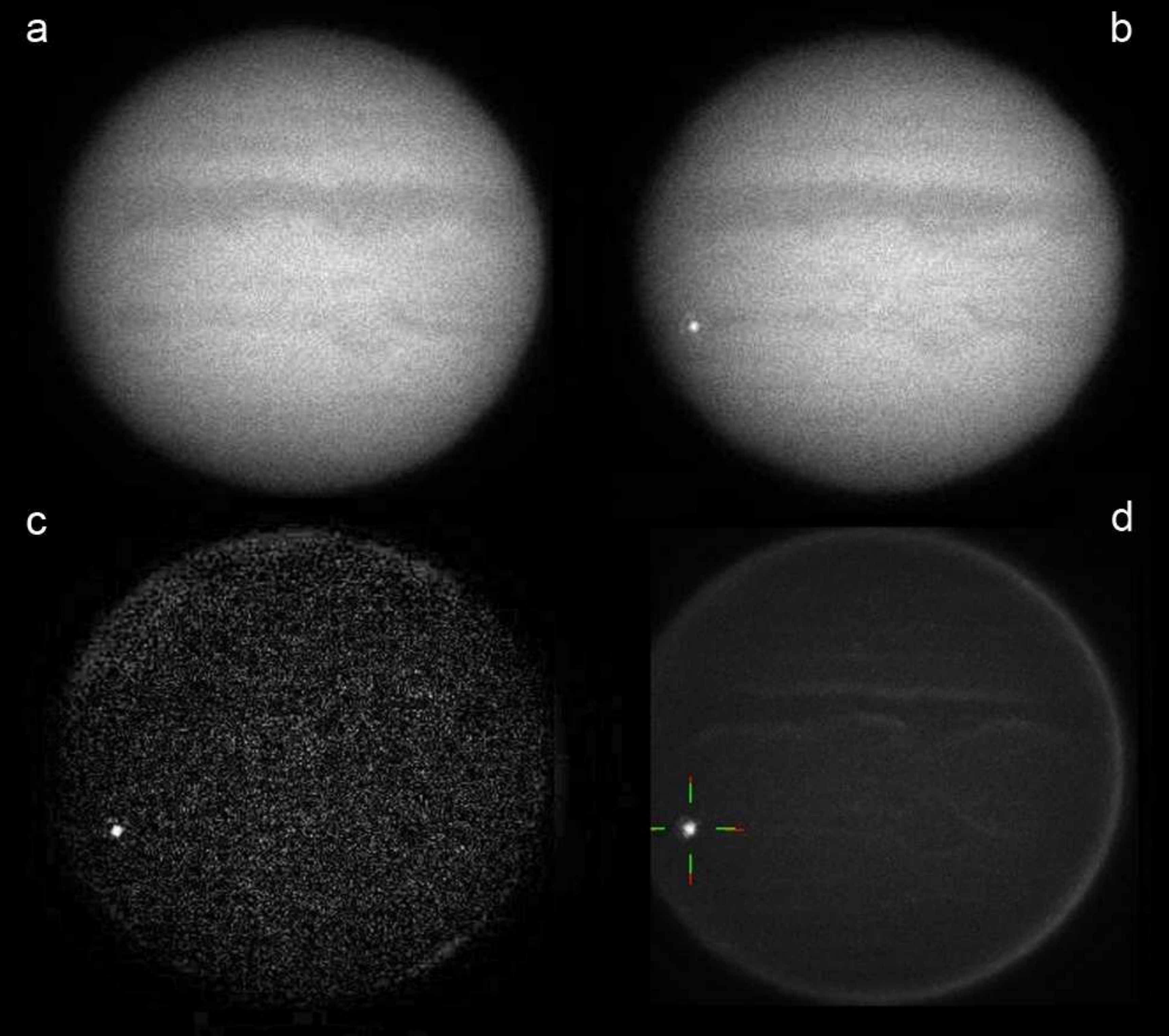}
    \caption{Raw frames of Jupiter and impact detection. (a) Raw frame before the start of the impact flash. (b) Brightest frame with the impact visible. (c) Differential image between b and a. (d) Detection image obtained with DeTeCt. This image shows the maximum brightness of a pixel minus the mean value and is constructed from the analysis of a full video of the planet.}
    \label{fig:obs}
\end{figure}

\subsection{Impact detection and raw lightcurve}
\label{sec:obs}
The last flashing impact on Jupiter was recorded on August 7th 2019 at 04:07:30 UTC by Ethan Chappel, an amateur observer who was observing Jupiter with an 8" aperture telescope and a ASI 290MM camera using a Chroma red filter and acquiring images at a rate of 83 frames per second (fps). The flash was captured by the video camera but not observed directly by E.C. on the screen. Instead, the flash appeared on a routine analysis of video observations using the software DeTeCt, which was designed to find short flashes of light on Jupiter video observations \citep{Hueso2018b}. DeTeCt coregisters image frames in a video observation correcting effects of atmospheric turbulence and performs an analysis of differential photometry of the video to find short-lived flashes producing a report of the analysis and a detection image (Figure~\ref{fig:obs}). Results from the use of DeTeCt by several observers will be discussed elsewhere.

\begin{figure}
    \centering
    \includegraphics[width=\columnwidth]{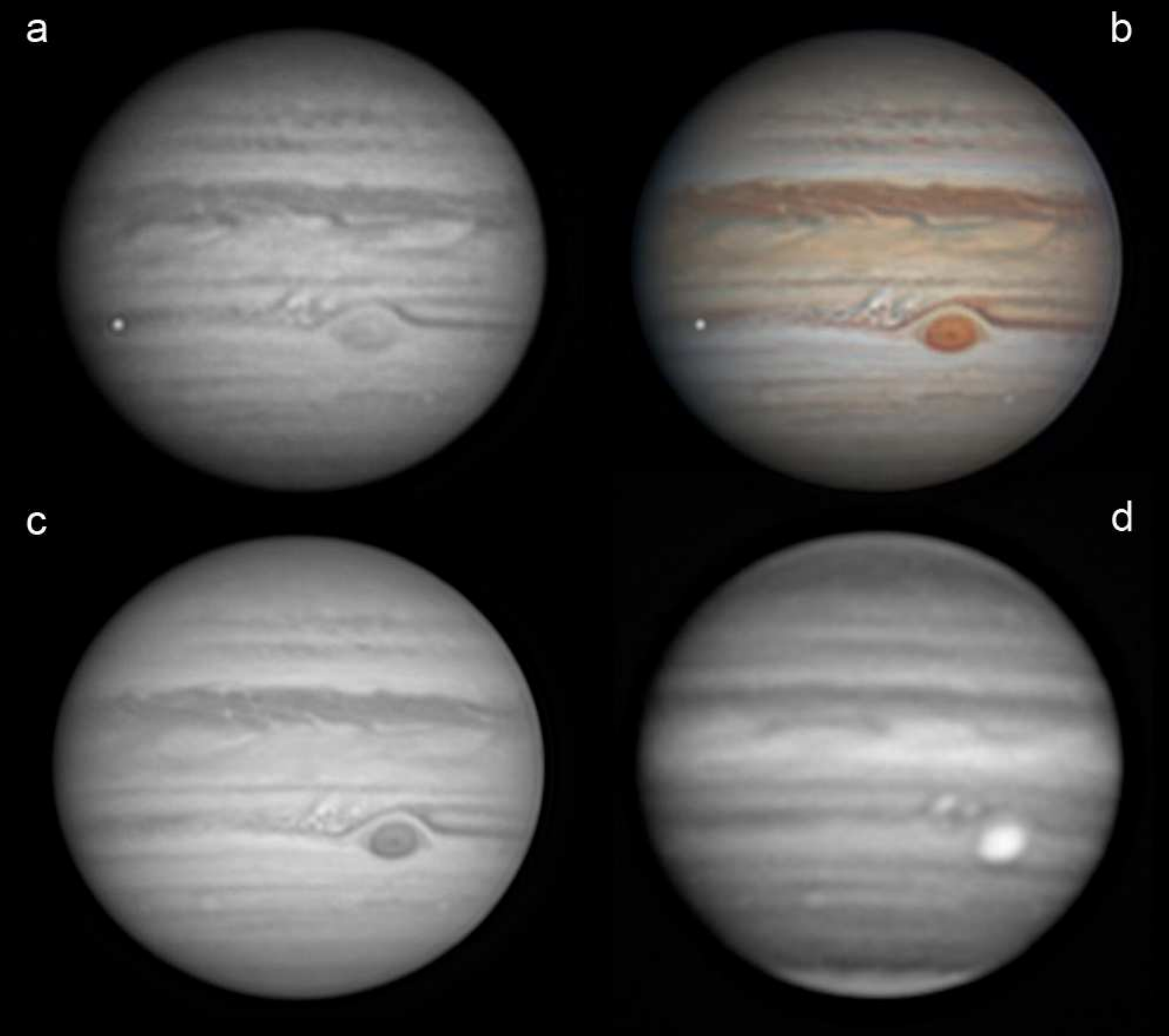}
    \caption{Impact flash on Jupiter on 2019-08-07 observed at 04:07:30 UT. (a) Image obtained from a stack of several thousand frames of the planet observed in the C8 Chroma filter. The flash has been processed separately from the few tens of frames where the impact is observable. (b) Image obtained composing the flash observations with previous and later videos of the planet in different filters and compensating the planetary rotation between the different videos with the software WinJupos. (c) Image of Jupiter based on a video obtained with a green filter 3 minutes after the impact. (d) Image of Jupiter based on a video obtained with a 890nm filter 28 minutes after the impact.}
    \label{fig:impact_location}
\end{figure}

The flash was located in the South Equatorial Belt at about 60$\degree$ W of the Great Red Spot at planetographic latitude of $19\degree$ S. Figure~\ref{fig:impact_location} shows a processed version of videos obtained close to the impact with the impact superimposed and based on the few dozens frames when the impact was visible and bright. Later observations recorded by E.C. on the same night, including an observation at the strong methane absorption band in 890nm, did not show any visible feature in the planet at the impact's location (Figure~\ref{fig:impact_location}). Later observations of the same area acquired by several observers on different wavelengths did not show an impact feature in the planet.

\begin{figure}
    \centering
    \includegraphics[width=0.85\columnwidth]{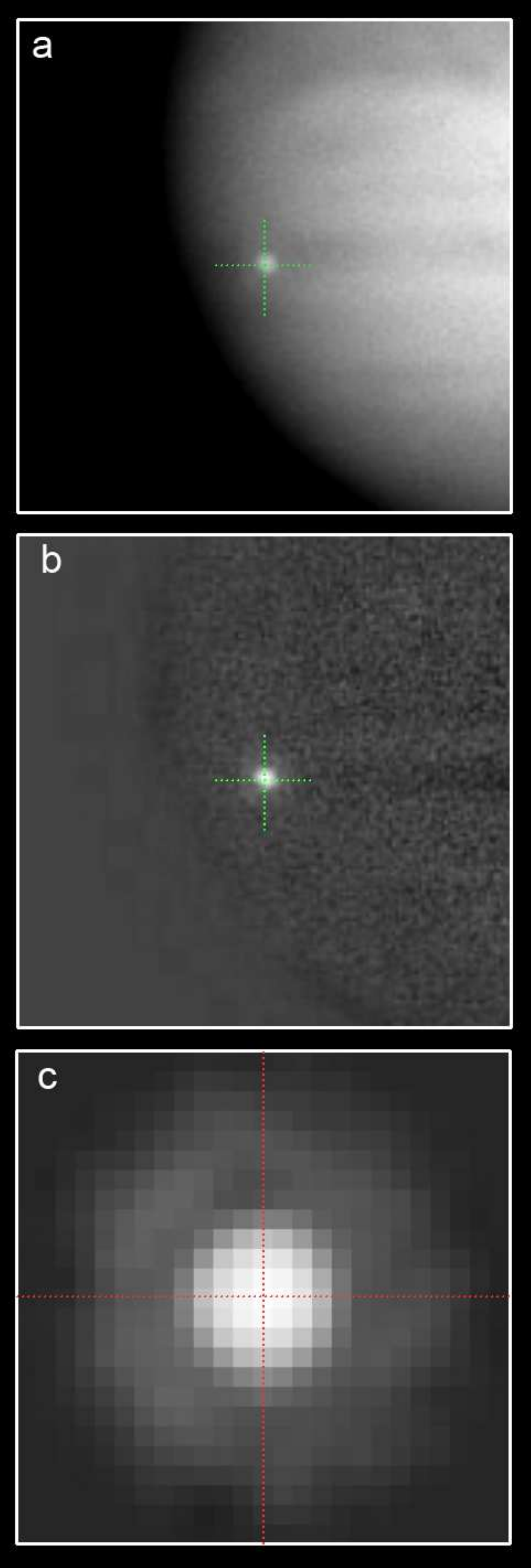}
    \caption{Detailed analysis of the spatial structure from the flash. (a) Stack of 5 images centered at the time of the brightest part of the flash. (b) Same image minus a reference image built from 5 images without the flash. All frames have been coregistered. (c) Zoom over the flash location showing the impact source and a diffraction ring around it. All the light in the central flash and diffraction ring is taken into account to build the lightcuve.}
    \label{fig:vert_structure}
\end{figure}

\begin{figure}
    \centering
    \includegraphics[width=\columnwidth]{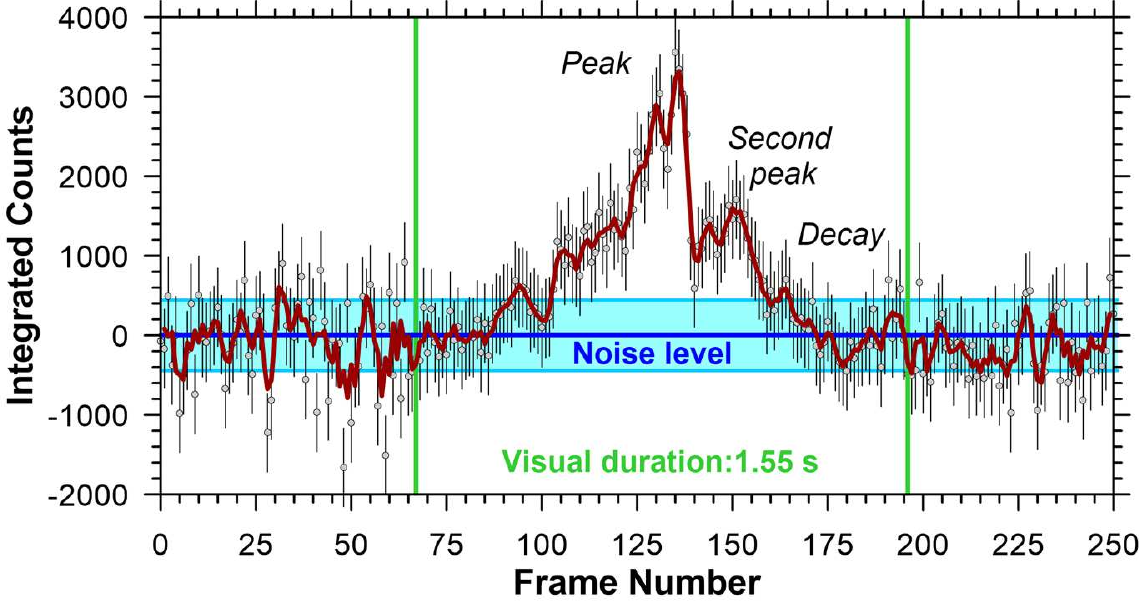}
    \caption{Impact light curve. Grey symbols represent the integrated light from the impact flash. Error bars are obtained by calculating the flash light with different parameters of the aperture mask. Red-curve is a running average of 3 frames. The visual duration of the flash is estimated from a careful visual inspection of all frames. The start of the flare is between frames 75-85, and the decay ends near frames 175-195. The noise level is estimated from the statistic of the light curve before and after the flash.}
    \label{fig:rawlc}
\end{figure}

We analyzed 250 frames of the red filter video with the impact with a pipeline that coregisters the individual frames and performs aperture photometry on the impact location. We tested different apertures and concluded that the best photometry was obtained with an aperture mask of 12 pix in radius and subtracting the background signal estimated from an outer annulus with inner radius of 14 pix and outer radius of 17 pix (Figure~\ref{fig:vert_structure}). The light-curve produced is shown in Figure~\ref{fig:rawlc} and contains significant structure with a central bright peak, a second peak and a decay. The flash was visible for a minimum total time of 1.16 and possibly 1.55 s when looking at subtle brigthness variations of the area. This range of durations is very similar to previous impacts in the planet \citep{Hueso2018a}. 

\subsection{Energy and mass estimate}
\label{sec:massest}
At the peak of its brightness, the impact was as bright as 0.082\% of Jupiter, equivalent to a +5.3 magnitude star, or as bright as Jupiter's moon Io. This brightness is similar to previous flash impacts on Jupiter but the degree of structure in the light-curve is higher allowing us to perform a more in depth analysis.

\begin{figure}
    \centering
    \includegraphics[width=\columnwidth]{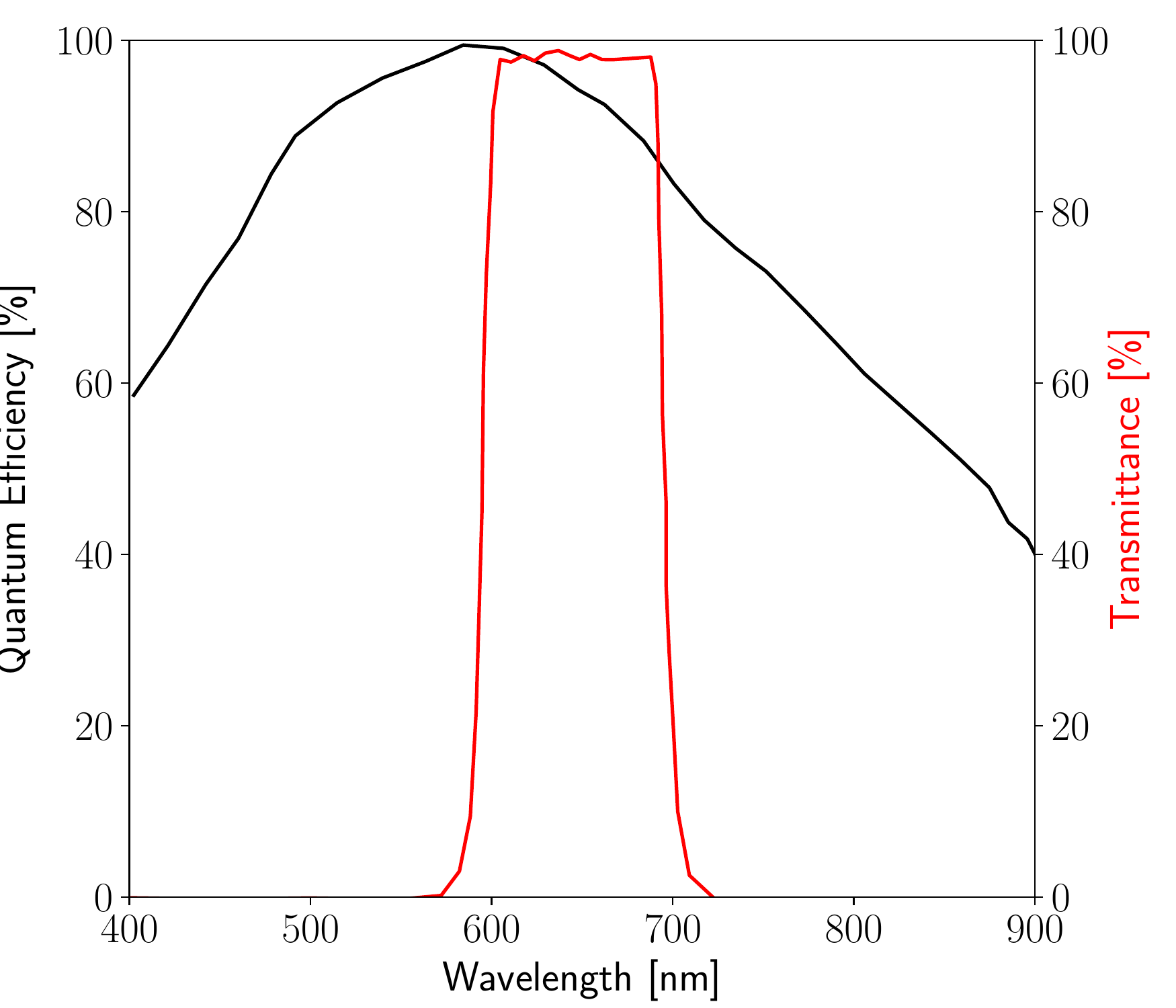}
    \caption{Quantum efficiency of the ZWO ASI290MM camera (black) and transmittance of the Chroma filter (red) used to obtain the lightcurve of the impact flash. }
    \label{fig:QE}
\end{figure}

To obtain the total released energy, we approximated the luminous energy in all wavelengths by assuming that the emitted light was blackbody radiation ($B_{\lambda}(T)$) between 3500-10000 K \citep{Hueso2013}. By integrating the lightcurve in the red filter, we obtain an energy of $E_\text{red} = 4.99 \times 10^{12}$ J, or about $1.19$ kt of TNT. This is converted to the luminous energy, assuming a blackbody emission of temperature $T$ using
\begin{equation}
    E_0 = E_\text{red} \dfrac{\int_{0}^{\infty} B_\lambda(T) d\lambda}{\int_{\lambda_1}^{\lambda_2} S_r(\lambda) B_\lambda(T) d\lambda},
\end{equation}
where $S_r(\lambda)$ is the wavelength-dependent camera quantum efficiency convolved with the filter transmittance (Fig~\ref{fig:QE}). To determine the total impact energy, we use the relation from \citet{brown2002nature}, given by
\begin{equation}
    E = 8.2508(E_0)^{0.885},
\end{equation}
where $E$ is the total kinetic energy and $E_0$ is the luminous energy, both in kt of TNT. Given the range of temperatures, the luminous energy of this bolide is between 15.9 kt to 26.7 kt. The corresponding total impact energy is between 96 to 151 kt. 
Due to the non-linearity of the conversion from filter to impact energy, the average energy cannot simply be determined as the average of this range. We assume that all blackbody temperatures in the given range are equally probable, and accordingly obtain a distribution of impact energies. For the remaining analysis, we use the mean of this distribution of $E=112 \pm 15$ kt.  Consequently, the lightcurve in the red filter is converted to total energy loss by multiplying by the factor, 
\begin{equation}
    f = \dfrac{E_\text{red}}{E} = 0.0106. 
\end{equation}

Assuming the impactor fell with a velocity between $60-70$ km/s, the corresponding mass of the object is between $190-260$ tonnes. 

\section{Fragmentation Model}
\label{sec:fragmodel}

To model the lightcurve, we implemented a fragmentation model based on \citet{Avramenko2014} and \citet{Wheeler2017}. In this model, the meteor is treated as single, fractured body, with an initial bulk strength $\sigma_0$. The speed $v$, mass $M$, height $h$ and angle $\theta$ with respect to the horizontal for the object's motion in the atmosphere is given by
\begin{eqnarray}
    \dfrac{dv}{dt} & = & -  \dfrac{C_D S \rho_a(h) v^2}{2 M} + g \sin(\theta), \\
    \dfrac{dM}{dt} & = & - \dfrac{ S \sigma_\text{ab} \rho_a(h) v^3}{2}, \\
    \dfrac{dh}{dt} & = & -v \sin(\theta), \\
    \dfrac{d\theta}{dt} & = & \dfrac{g\cos(\theta)}{v} + \dfrac{v\cos(\theta)}{R_p + h},
\end{eqnarray}
where $S$ is the cross-section area, $C_D$ is the drag coefficient and $\sigma_\text{ab}$ is the ablation coefficient (the amount of evaporated material per unit energy in kg/J). $\rho_a$ is the atmospheric density as a function of height. 

When the ram pressure, 
\begin{equation}
    P_\text{ram} = \rho_a v^2
\end{equation}
exceeds the strength of the body ($\sigma$), it is fragmented into a number of equal-sized objects which travel together. The smaller bodies are less fractured and assumed to have a higher strength given by a Weibull-like relation \citep{Weibull1951},
\begin{equation}
    \sigma = \sigma_0\left(\dfrac{M}{M_\text{fr}}\right)^\alpha 
\end{equation}
where $M_0$ is the total mass of the body and $M_\text{fr}$ is the mass of each fragment. The number of equal-sized fragments is given by
\begin{equation}
    N_\text{fr} = \dfrac{16 S^3 \rho_m}{9 \pi M^2}
\end{equation}
and the mass of each fragment is
\begin{equation}
    M_\text{fr} = \dfrac{M}{N_\text{fr}} = \dfrac{9 \pi M^3}{16 S^3 \rho_m}
\end{equation}
where $\rho_m$ is the bulk density of the main body. 

The cross-sectional area of the bolide can decrease as mass is ablated, or increase as the object fragments. Fragmentation occurs when the object is experiencing pressure higher than its bulk strength. Therefore, the rate at which the cross-sectional area changes is given by
\begin{equation}
    \dfrac{dS}{dt} = 
        \begin{cases}
            \dfrac{2}{3}\dfrac{S}{M}\dfrac{dM}{dt} & P_\text{ram} < \sigma, \\
            \dfrac{2}{3}\dfrac{S}{M}\dfrac{dM}{dt} + C_\text{fr} \dfrac{\left(P_\text{ram} - \sigma\right)^{1/2} S}{M^{1/3} \rho_m^{1/6}} & P_\text{ram} > \sigma,
        \end{cases}
\end{equation}
where $C_\text{fr}$ is a dimensionless parameter describing how tightly held the fragments are.

\begin{figure}
    \centering
    \includegraphics[width=\columnwidth]{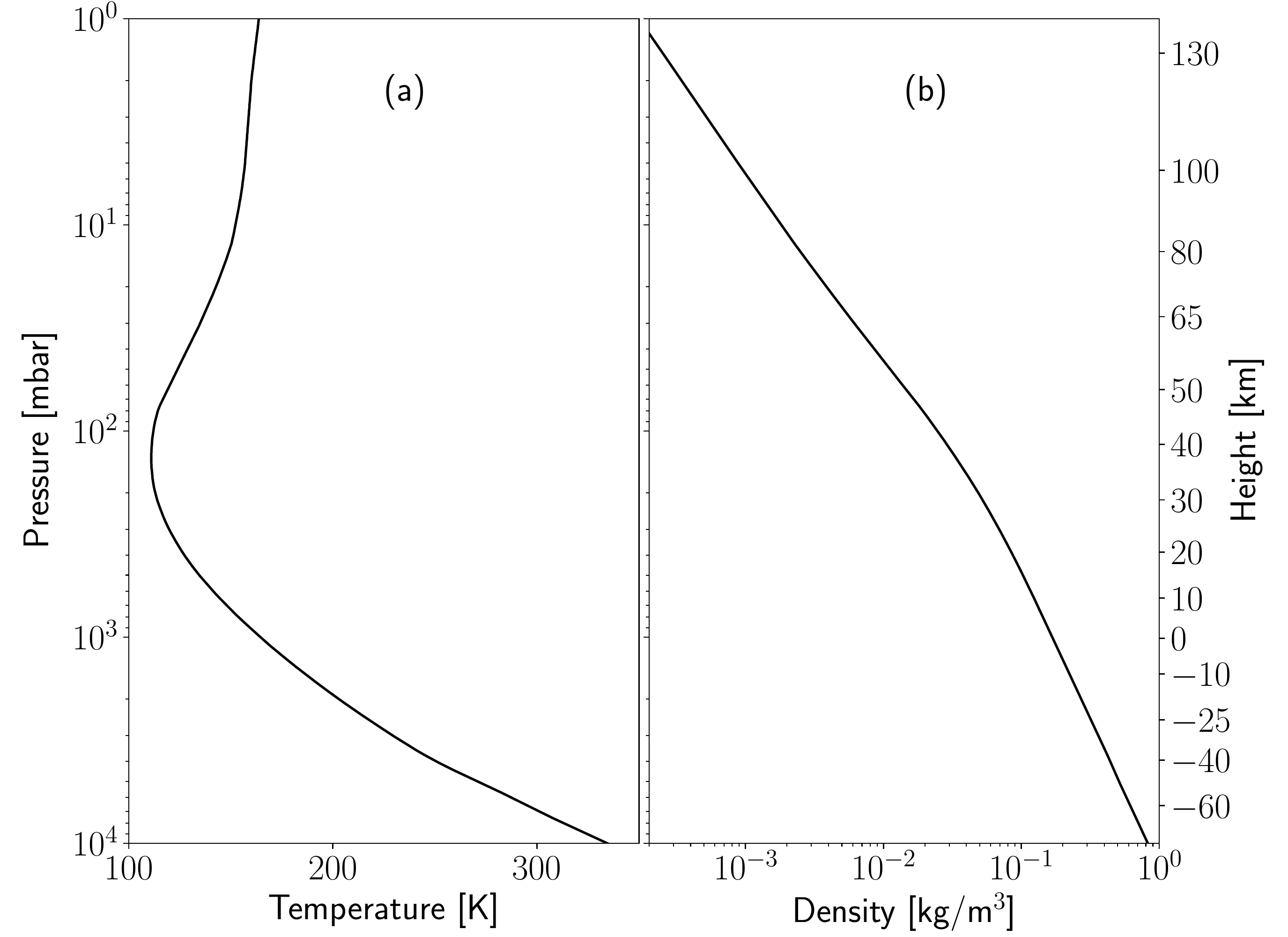}
    \caption{Temperature pressure profile (a) and corresponding density profile (b) of Jupiter's atmosphere used as input to the fragmentation model. }
    \label{fig:tpprof}
\end{figure}

We use $C_D = 0.92$ \citep{CarterJandirKress2009}, $g=24.0$ m/s, $R_p=70000$ km as fixed constants for all test cases. 
The atmospheric density profile was constructed by solving the hydrostatic balance given the temperature-pressure profile adapted from \citet{DemingHarrington} (Fig~\ref{fig:tpprof}a) and assuming a dry atmosphere with a constant molar mass of $M_\text{atmo} = 2.28$. The corresponding density profile is shown in Fig~\ref{fig:tpprof}b. By convention, $h=0$ corresponds to the 1 bar pressure level. 

Aside from fragmentation described above, we also allow for the main body to break up into several discrete fragments, which are modelled as independent objects under the same principles, similar to \citet{Palotai2019}. We prescribe the mass of the fragment, location (i.e. dynamical pressure) of each fragmentation event and $C_\text{fr}$ of the new fragments. If needed, the new fragments can also have a different strength-scaling relation $\alpha$ and bulk strength $\sigma_0$. Therefore, in total, we have the following free parameters: initial velocity, flight angle, $\alpha$, $\sigma_0$ and $\rho_m$ of the main body. The initial mass is calculated from energy conservation, i.e.
\begin{equation}
    \label{eq:initmass}
    M_0 = \dfrac{2 E}{v_0^2}.
\end{equation}
For each fragmentation event, we have $M$, $P_\text{release}$, $C_\text{fr}$ and optionally $\sigma_0$ and $\alpha$. These parameters are determined by doing a trial-and-error fit of the lightcurve. To simplify the parameter space, we assume that these independent objects only break from the main body; they do not further disintegrate into independent bodies. 

\section{Test cases}

\begin{table}
    \centering
    \begin{tabular}{|c|c|c|c|c|}
        \hline
        \multirow{2}{*}{Case} & $\rho_m$   & Speed  & Mass  & Diameter \\
              & [kg/m$^3$] & [km/s] & [ton] & [m] \\
        \hline
        \hline
         1 & 500 & 60 & 262 & 10.0 \\
         2 & 2500 & 60 & 262 & 5.85 \\
         3 & 5000 & 60 & 262 & 4.64 \\
         4 & 500 & 65 & 223 & 9.48 \\
         5 & 2500 & 65 & 223 & 5.54\\
         6 & 5000 & 65 & 223 & 4.40\\
         7 & 500 & 70 & 192 & 9.02\\
         8 & 2500 & 70 & 192 & 5.27\\
         9 & 5000 & 70 & 192 & 4.19\\
         \hline
    \end{tabular}
    \caption{Input parameter space for the different cases tested}
    \label{tab:testcases}
\end{table}

On Earth, the trajectory of the bolide can be determined using video cameras that record the impact from different veiwpoints. The trajectory is used to calculate an energy-deposition profile as a function of height. The trajectory defines the impact geometry (initial velocity and angle), while the energy deposition profile helps constrain the material properties (material strength/density). For jovian impactors, it is not possible to obtain these initial parameters. Therefore, in our case, we test three separate types of objects: cometary ($\rho_m=500$ kg/m$^3$), chondritic ($\rho_m=2500$ kg/m$^3$) and iron-nickel ($\rho_m=5000$ kg/m$^3$). For each case, we test three initial velocities: $v=60$ km/s, $65$ km/s and $70$ km/s, which covers the range of impactor velocities on Jupiter.
For each velocity, we compute the initial mass from (\ref{eq:initmass}) and diameter from the assumption of a perfect sphere. A list of the test cases is shown in Table~\ref{tab:testcases}. For different material types, the key difference is the bulk strength ($\sigma_0$) and the ablation coefficient ($\sigma_\text{ab}$). In general, a lower bulk strength causes the meteor to be disrupted higher up, and thus flare for longer (i.e. the peaks are wider). A high strength meteor has a narrow and tall peak flare. To reduce the number of free parameters for the different cases, we fix these two values for each material type and assume they do not change during breakup. Inhomogeneities in the object are thus represented by discrete fragments rather than during disruption.

\subsection{Cometary}
The cometary case has a low initial bulk strength and a high $\sigma_\text{ab}$. We use an initial bulk strength of $10$ kPa \citep{Trigo2006} and an ablation coefficient of $\sigma_{\text{ab}}=2\times10^{-8}$ kg/J \citep[Type A from][]{Ceplecha1988}. A higher ablation coefficient results in the meteor evaporating before reaching the first peak. 

\subsection{Stony}
For the stony cases, bulk strengths are a strong function of porosity and structure, varying between $\sim0.1-1$ MPa \citep{popova2011mps}, with more massive meteors having a lower strength. Since it is difficult to effectively test the differences between porosities, we use a fixed value of $0.5$ MPa for stony meteorites. For the ablation, we use a value of $\sigma_{\text{ab}}=2\times10^{-9}$ kg/J, corresponding to an ordinary chondritic meteor from \citet{Ceplecha1988}.

\subsection{Iron-nickel}
Due to the low number of studied iron fireballs, the bulk strength is poorly constrained. The compressible strength tested using the recovered fragments can reach $\sim 100$ MPa \citep{Chyba1993}, but this is usually orders of magnitude greater than the bulk strength for the entire body due to the presence of cracks/fractures. We expect a tightly packed iron meteor to have a bulk strength higher than that of the stony case. Above $\sim 5$ MPa bulk strength, disruption occurs only near the main peak, which makes it difficult to create the earlier flares. At a bulk strength of about $2$ MPa, disruption happens at between $t \sim -0.4$s, which is consistent with the main flare. Therefore, for our iron cases, we use $\sigma_0 = 2$ MPa.

On the other hand, ablation is much more prominent for iron meteors since the high thermal conductivity allows the higher temperatures to penetrate the interior, causing faster melting. \citet{Revelle1994} find that the ablation coefficient for known iron meteors is between $\sigma_\text{ab} \sim 10^{-9} -  10^{-7}$ kg/J from 6 observed fireballs. The lower end is similar to the value used for stony meteroids. Above $\sigma_{\text{ab}}\sim 2\times10^{-8}$ kg/J, the ablation is too strong to produce the last two peaks. Therefore, for the iron case, we use $\sigma_\text{ab} = 10^{-8}$ kg/J, which is in the middle of the range determined by \citet{Revelle1994}.

For these nine test cases, we vary all other parameters to match the lightcurve as close as possible. To quantify the results in each, we calculate the average absolute residual and the difference in peak energy release between the modelled and observed lightcurves.
\section{Results and Discussion}
\label{sec:results}

\begin{figure*}
    \centering
    \includegraphics[width=\textwidth]{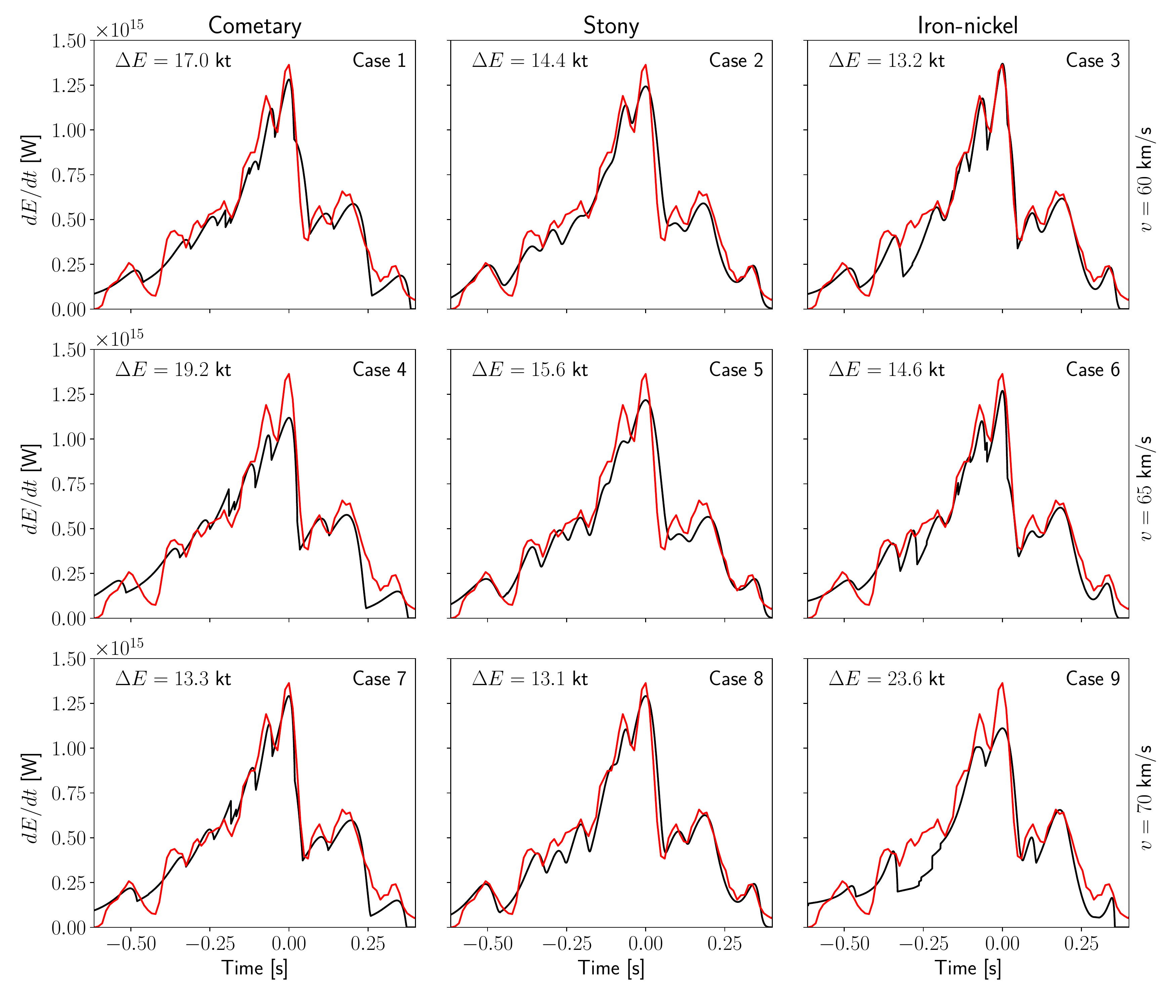}
    \caption{Fragmentation model output (black) compared to observed lightcurve (red) for each case. Case numbers are shown in the upper right hand corner as defined in Table~\ref{tab:testcases}. We calculate the average residual of each case from the observed lightcurve, as defined in Table~\ref{tab:bodyparams}. }
    \label{fig:alllc}
\end{figure*}

\begin{table*}
\begin{tabular}{|c|c|c|c|c|c|c|c|c|c|c|}
\hline
Case & Material & $\theta$    & $\alpha$ & $C_\text{fr}$ & Peak height      & Peak pressure      & Residual in peak          & End height       & End pressure      & Avg. Residual    \\
     &          &             &          &               & {[}km{]} & {[}hPa{]} & {[}$10^{13}$ W{]} & {[}km{]} & {[}hPa{]} & {[}kt{]} \\ \hline
1    & C        & $65\degree$ & 0        & 1.3           & 147              & 0.71               & $-8.20$                   & 126              & 1.73              & 17.0             \\ \hline
2    & S        & $65\degree$ & 0.03     & 1.5           & 103              & 4.85               & $-12.4$                   & 78               & 14.4              & 14.4             \\ \hline
3    & IN       & $65\degree$ & 0.03     & 1.1           & 86               & 10.2               & $0.592$                   & 67               & 26.3              & 13.2             \\ \hline
4    & C        & $50\degree$ & 0.01     & 1.3           & 152              & 0.57               & $-24.5$                   & 134              & 1.26              & 19.2             \\ \hline
5    & S        & $50\degree$ & 0.03     & 1.5           & 118              & 2.50               & $-14.6$                   & 96               & 6.46              & 15.6             \\ \hline
6    & IN       & $50\degree$ & 0.05     & 1.1           & 91               & 8.29               & $-9.44$                   & 73               & 19.1              & 14.6             \\ \hline
7    & C        & $50\degree$ & 0.01     & 1.3           & 155              & 0.50               & $-7.24$                   & 136              & 1.16              & 13.3             \\ \hline
8    & S        & $50\degree$ & 0.03     & 1.5           & 113              & 3.14               & $-7.28$                   & 90               & 8.36              & 13.1             \\ \hline
9    & IN       & $45\degree$ & 0.05     & 1.1           & 82               & 12.3               & $-25.3$                   & 65               & 28.5              & 23.6             \\ \hline
\end{tabular}
\caption{Results for the nine cases. The material type of each case is listed (C=cometary, S=stony, IN=iron-nickel). $\theta$ (entry angle with respect to the horizontal), $\alpha$ and $C_\text{fr}$ define the initial parameters of the main body. We calculate the location (altitude and atmospheric pressure) and compare its strength to the observed peak energy release of $1.36\times10^{15}$ W. The end height is the final height of all remaining fragments and the terminal pressure is the atmospheric pressure at this altitude. The average residual is the mean of the absolute difference between the modelled and observed energy deposition profile.}
\label{tab:bodyparams}
\end{table*}

The results from the nine cases are shown in Fig~\ref{fig:alllc}. The input parameters for the main body for the different cases is shown in Table~\ref{tab:bodyparams}. The individual fragmentation points for each case are detailed in Tables~\ref{tab:case1}-\ref{tab:case9}. The main flare is best matched by Case 3 (iron-nickel at $60$ km/s), while the lowest residual is for Case 8 (stony at $70$km/s), although the residuals for Cases 7 and 3 are similar. For the slow cases, the flares for the weaker material is too wide, as the material ablates for longer, and higher up. The iron case is able to penetrate into greater depths (reaching terminal altitudes of $\sim60-70$km), resulting in a sharp disintegration as the bolide reaches the denser atmosphere. For the fast cases, the cometary case is able to reproduce the shape of the peaks, but the iron case burns too quickly to produce the necessary energy at depth to match the peak intensity. 

\begin{figure*}
    \centering
    \includegraphics[width=\textwidth]{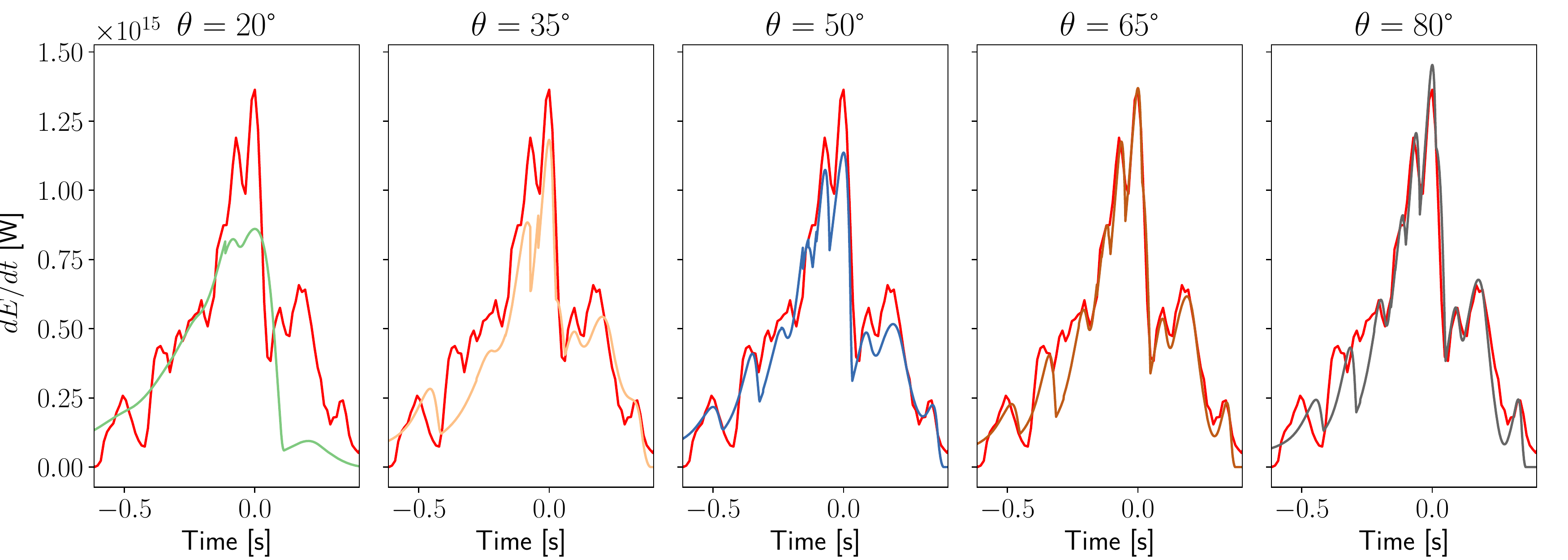}
    \caption{Fits for Case 3 (60 km/s iron-nickel) as a function of entry angle. All other main body parameters are kept the same as defined in Table~\ref{tab:bodyparams}. Slight changes to $P_{\text{release}}$  and $\sigma_0$ (Table~\ref{tab:case3}) for the individual fragmentation events were needed to match the lightcurve. For the $\theta=20\degree$ case, it was not possible to produce the last three peaks because there was insufficient mass after the main flare. }
    \label{fig:angle_lc}
\end{figure*}

The fits for different entry angles of the bolide for the 60 km/s iron-nickel material (Case 3) is shown in Figure~\ref{fig:angle_lc}. In this case, the $\theta=65\degree$ case best matched both the strength and the width of the peaks, and was chosen as the most likely scenario. A steep entry angle caused the impactor to reach the deeper atmosphere quickly and thus resulted in a sharper and taller peaks. In the $\theta=80\degree$ case, it can be seen that the shapes of the main peaks are well matched, but the width of the last two are slimmer than observed. Contrarily, a shallow entry angle caused the meteor to ablate longer before reaching the altitude of disruption. This made it difficult to reproduce the energy release from the later flares, as is evident in the $\theta=20\degree$ case where there was insufficient mass to reproduce the flares. The method to determine the parameters for the other cases were done with a similar fashion.

The cometary case required various inhomogeneities in the material makeup to match the observed lightcurve. For all velocities, the smooth bump at $-0.2$s, $-0.1$s and the peak at $0.2$s required higher strength scaling coefficients compared to the main object. Decreasing the strength scaling of these peaks caused them to be thinner than observed and also shift earlier in time, due to the fragments having a smaller material strength. Furthermore, in all the cometary cases, the drop in brightness at $-0.45$s is not resolved. The last two peaks at $0.2-0.4$s are also non-symmetric and burn slowly to a sharp drop compared to the more gaussian-like appearance of the observed lightcurve. A lower ablation rate is necessary to match these features. For the 70 km/s case, the main flare with its two peaks are matched well, although the last peak during the decay is much lower than observed. The cometary meteors peak at an altitude of $\sim140-150$ km, at an atmospheric pressure of $\sim0.5$ hPa and reach a terminal depth of $\sim120-130$ km (pressure of $\sim1-2$ hPa). 

For all three velocities, the stony meteor produces essentially the median result: the lightcurve is followed for the most part, but the peaks are too wide and fall short of the observed energy release. However, all three stony cases are better matches to the earlier part of the lightcurve (between $t\sim-0.6$ to $-0.2$s), reproducing both the drop at $-0.45$s and the three successive flares, especially in the $65$ km/s case. The 70 km/s stony case (Case 8) has the smallest residual, but falls short on both the main and secondary peaks. The main peak is also wider than observed, although the last three peaks are well modelled. A lower ablation coefficient increases the strength but widens the shape of the peaks. The stony meteors reach much lower depths compared to the cometary impactors, peaking at an altitude of about $100-120$ km ($P\sim3-10$ hPa) and have a terminal depth of about $80-90$ km ($P\sim10-20$ hPa). 

The 60km/s iron meteor (Case 3) produces the best match of the main peak and decay. Increasing the velocity, however, reduces the goodness of fit. The 70km/s case ablates too quickly early on to maintain the required energy for the main peak. The flares before $\sim0.2$s are poorly modelled. It is likely that the outer layers of the object, which contribute to the earlier part of the lightcurve, are weaker than modelled here and are similar to the eroding dust grains \citep{Borovicka2007,Borovicka2013}. These dust grains produce a gradual increase and decrease in brightness (a ``hump") and are followed by a deceleration of the main body. Due to the lack of direct observational constraints on the velocity, it is not possible to model such fragments well. As expected, the iron meteors reach the deepest altitudes, peaking at about $80-90$ km ($P\sim10$ hPa) and terminating at about $60-70$ km ($P\sim20-30$ hPa). 

Regardless of composition and velocity, the first two peaks at $t=-0.5$s and $t=-0.4$s are modelled by weak fragments (on the order of the initial strength of the main body). The release of these fragments occurs before the main body is disrupted for the stony and iron-nickel meteors. For the cometary cases, these fragments have a material strength similar to the dynamic pressure experienced by the bolide when the release occurs. 

Conversely, the last three peaks after the main flare require fragments that are stronger than the main body. We interpret these fragments to consititute the ``nucleus" of the body where the grains are packed much tighter than the outer layers and fractures are minimal. In some cases the ratio of the fragment strength to main body was more than an order of magnitude (Cases 1, 2, 4, 7, 9), while in others, the ratio was lower (Cases 3, 5, 6). Such large gradients in material strength have been observed on bolides on Earth \citep{Avramenko2014,Wheeler2018}.

These cases are sensitive to material strength and ablation coefficient, and a further study of these values are necessary to refine our model parameters, especially for the cometary and metallic cases. Without the data on energy deposition at different heights, however, it is difficult to successfully distinguish small differences in material makeup. Further parameterization is likely needed to explain the differences seen here, such as the addition of chain fragmenting for child fragments. Implementing a self-consistent modelling framework for fragmentation is necessary before the parameter space is expanded, but this is challenging given the trial-and-error nature of the analysis. 

Furthermore, the higher velocity cases required a shallower entry angle to better fit the lightcurve. This results in the peak and end height of all three velocity cases being roughly similar (to within $\sim10$ km), probably a consequence of the assumption of constant material strength and ablation coefficient for each material type. Impact geometry can have large effects on terminal depth \citep{Pond2012}, and requires further investigation, especially for jovian impactors. In our model, terminal depth is a strong function of discrete fragmentation events, thus making a direct relation between impact geometry and altitudes difficult. A more rigorous study of the parameter space is necessary to probe this relation.

Nevertheless, we find several distinguishing features between different materials for a given entry velocity and angle. Cometary meteors ablate heavily on the leading edge compared to the trailing edge, making the flares less symmetric. Ablation and disruption also occurs higher up for these objects given their low bulk strength, making the lightcurve from a cometary impactor brighter than observed before the main flare. 

Chondritic and iron meteors produce sharper and slimmer peaks due to penetrating deeper into the atmosphere before being disrupted. The shapes of the peaks are, in most cases, better fits to the observation with less inhomogeneities compared to the cometary cases. Meteor disruption for these materials also happens at the start of the main flare, as opposed to before the initial flare. Consequently, it is more probable that the impactor was chondritic/metallic rather than cometary. For the slower cases, a higher ablation coefficient (i.e. metallic object) is a better match. For the high velocity case, a lower ablation rate (i.e. stony material) produces much better fits, as a strong, but highly ablating meteor loses too much mass before its first peak. The flares before $-0.2$s are best modelled by a weak outer layer and the last peaks during the decay are indicative of strong nucleus.

\section{Conclusion}
We analysed an observation of an impact flash on Jupiter by Ethan Chappel on August 7th 2019 and obtained the lightcurve using differential photometry of the impact area over coregistered images. The total energy release from the impact was between $96 - 151$ kt depending on the assumption of luminous efficiency, which corresponds to a mass between $190-260$ metric tonnes. We developed a fragmentation model where we input the physical parameters of the bolide such as entry velocity, angle and material makeup to model the energy release from the bolide. We considered three materials (cometary, stony and iron) at three velocities (60, 65 and 70 km/s) and modified the parameters of the model using a trial-and-error to match the lightcurve.

We find that the most likely scenarios are a strong, slow bolide, or a fast, weak impactor made of a combination of stony and iron material. For intermediate velocities, the stronger material is preferred to produce both the shape and intensity of the discrete fragmentation peaks. The initial flares are best reproduced by material that is weaker than required to produce the main peak, such as an envelope of dust grains. The last three peaks in the lightcurve require very strong material, likely corresponding to the densest parts of the object. The impactor was too small to leave a debris field in the jovian atmosphere. 

This impact event was recorded with an unprecedented quality that allowed the detailed modelling presented in this paper. We expect that future detections of impacts will be done with similar fast high-quality cameras allowing for a partial determination of the physical characteristics of small objects impacting Jupiter.

\section*{Acknowledgements}
We are very grateful to the ensemble of amateur astronomers running DeTeCt on their video observations of Jupiter. R.H. and A.S.L. were supported by the Spanish project AYA2015-65041 (MINECO/FEDER, UE) and Grupos Gobierno Vasco IT1366-19. DeTeCt has been partially supported by the Europlanet 2020 Research Infrastructure. Europlanet 2020 RI has received funding from the European Union's Horizon 2020 research and innovation programme under grant agreement No 654208. We also thank the reviewer whose comments have improved the clarity of the manuscript. 




\bibliographystyle{mnras}
\bibliography{ref} 




\appendix

\section{Fragmentation events}

The discrete fragmentation parameters for each case are shown below in Tables~\ref{tab:case1}-\ref{tab:case9}.

\begin{table}
    \centering
    \begin{tabular}{|c|c|c|c|c|}
        \hline
        Mass & $P_\text{release}$ & $C_\text{fr}$ & $\alpha$ & $\sigma_0$ \\
        $\text{[ton]}$ & [MPa] & & & [MPa] \\
        \hline
        \hline
        5.20  & 0.05 & 3.0 & 0.00 & 0.06 \\
        5.20  & 0.06 & 3.0 & 0.00 & 0.10 \\
        5.20  & 0.15 & 2.0 & 0.06 & - \\
        7.80  & 0.21 & 2.0 & 0.00 & - \\
        10.39 & 0.32 & 3.0 & 0.02 & - \\
        7.80  & 0.28 & 3.0 & 0.06 & - \\
        15.59 & 0.27 & 1.2 & 0.00 & 0.12 \\
        57.17 & 0.27 & 1.8 & 0.06 & 0.20 \\
        15.59 & 0.28 & 1.8 & 0.01 & 0.58 \\
        \hline
    \end{tabular}
    \caption{Fragmentation events for Case 1.}
    \label{tab:case1}
\end{table}

\begin{table}
    \centering
    \begin{tabular}{|c|c|c|c|c|}
        \hline
        Mass & $P_\text{release}$ & $C_\text{fr}$ & $\alpha$ & $\sigma_0$ \\
        $\text{[ton]}$ & [MPa] & & & [MPa] \\
        \hline
        \hline
        12.99 & 0.10 & 2.0 & 0.01 & 0.25 \\
        10.39 & 0.10 & 2.1 & 0.01 & 0.45 \\
        10.39 & 0.70 & 2.2 & 0.00 & - \\
        20.79 & 0.75 & 3.0 & 0.05 & - \\
        12.99 & 1.10 & 1.8 & 0.03 & - \\
        12.99 & 1.60 & 2.0 & 0.00 & - \\
        15.59 & 1.20 & 1.1 & 0.00 & 1.40 \\
        54.57 & 1.20 & 2.0 & 0.06 & 1.55 \\
        10.39 & 1.20 & 2.0 & 0.01 & 4.50 \\
        \hline
    \end{tabular}
    \caption{Fragmentation events for Case 2.}
    \label{tab:case2}
\end{table}

\begin{table}
    \centering
    \begin{tabular}{|c|c|c|c|c|}
        \hline
        Mass & $P_\text{release}$ & $C_\text{fr}$ & $\alpha$ & $\sigma_0$ \\
        $\text{[ton]}$ & [MPa] & & & [MPa] \\
        \hline
        \hline
        7.80 & 0.70 & 2.0 & 0.03 & 0.90 \\
        12.99 & 0.90 & 2.5 & 0.02 & 1.50 \\
        12.99 & 0.90 & 3.0 & 0.03 & - \\
        10.39 & 3.20 & 1.8 & 0.02 & - \\
        18.19 & 3.70 & 1.6 & 0.00 & - \\
        10.39 & 4.90 & 1.5 & 0.00 & - \\
        12.99 & 4.40 & 2.4 & 0.03 & 5.30 \\
        54.57 & 4.40 & 2.2 & 0.07 & 5.30 \\
        15.59 & 4.60 & 2.4 & 0.02 & 12.50 \\
        \hline
    \end{tabular}
    \caption{Fragmentation events for Case 3.}
    \label{tab:case3}
\end{table}

\begin{table}
    \centering
    \begin{tabular}{|c|c|c|c|c|}
        \hline
        Mass & $P_\text{release}$ & $C_\text{fr}$ & $\alpha$ & $\sigma_0$ \\
        $\text{[ton]}$ & [MPa] & & & [MPa] \\
        \hline
        \hline
        4.43  & 0.05 & 3.0 & 0.00 & 0.06 \\
        4.43  & 0.06 & 3.0 & 0.00 & 0.10 \\
        4.43  & 0.15 & 2.0 & 0.06 & - \\
        6.64  & 0.21 & 2.0 & 0.00 & - \\
        8.86  & 0.32 & 3.0 & 0.01 & - \\
        4.43  & 0.28 & 3.0 & 0.04 & - \\
        13.29 & 0.27 & 1.2 & 0.00 & 0.08 \\
        48.71 & 0.27 & 1.8 & 0.06 & 0.17 \\
        11.07 & 0.28 & 1.8 & 0.01 & 0.52 \\
        \hline
    \end{tabular}
    \caption{Fragmentation events for Case 4.}
    \label{tab:case4}
\end{table}

\begin{table}
    \centering
    \begin{tabular}{|c|c|c|c|c|}
        \hline
        Mass & $P_\text{release}$ & $C_\text{fr}$ & $\alpha$ & $\sigma_0$ \\
        $\text{[ton]}$ & [MPa] & & & [MPa] \\
        \hline
        \hline
        11.07 & 0.05 & 2.0 & 0.01 & 0.15 \\
        8.86  & 0.45 & 2.1 & 0.00 & 0.02 \\
        11.07 & 0.45 & 2.0 & 0.00 & - \\
        13.29 & 0.55 & 2.0 & 0.00 & - \\
        4.43  & 0.85 & 1.8 & 0.00 & - \\
        8.86  & 0.85 & 1.8 & 0.03 & - \\
        14.39 & 0.70 & 1.1 & 0.00 & 0.70 \\
        46.50 & 0.70 & 2.0 & 0.06 & 0.80 \\
        8.86  & 0.70 & 2.0 & 0.01 & 2.30 \\
        \hline
    \end{tabular}
    \caption{Fragmentation events for Case 5.}    
    \label{tab:case5}
\end{table}

\begin{table}
    \centering
    \begin{tabular}{|c|c|c|c|c|}
        \hline
        Mass & $P_\text{release}$ & $C_\text{fr}$ & $\alpha$ & $\sigma_0$ \\
        $\text{[ton]}$ & [MPa] & & & [MPa] \\
        \hline
        \hline
        4.43  & 0.80 & 2.0 & 0.03 & 1.10 \\
        11.07 & 1.50 & 2.0 & 0.02 & 1.50 \\
        6.64  & 0.70 & 3.2 & 0.00 & - \\
        11.07 & 1.20 & 2.8 & 0.03 & - \\
        6.64  & 3.60 & 2.8 & 0.03 & - \\
        8.86  & 3.60 & 1.8 & 0.05 & - \\
        6.64  & 4.10 & 1.8 & 0.00 & - \\
        11.07 & 4.90 & 2.0 & 0.00 & - \\
        8.86  & 4.50 & 2.4 & 0.03 & 5.10 \\
        46.50 & 4.50 & 2.2 & 0.06 & 5.10 \\
        11.07 & 5.50 & 2.4 & 0.02 & 10.40 \\
        \hline
    \end{tabular}
    \caption{Fragmentation events for Case 6.}    
    \label{tab:case6}
\end{table}

\begin{table}
    \centering
    \begin{tabular}{|c|c|c|c|c|}
        \hline
        Mass & $P_\text{release}$ & $C_\text{fr}$ & $\alpha$ & $\sigma_0$ \\
        $\text{[ton]}$ & [MPa] & & & [MPa] \\
        \hline
        \hline
        3.82  & 0.05 & 3.0 & 0.00 & 0.06 \\
        3.82  & 0.06 & 3.0 & 0.00 & 0.10 \\
        3.82  & 0.15 & 2.0 & 0.06 & - \\
        5.73  & 0.21 & 2.0 & 0.00 & - \\
        7.64  & 0.32 & 3.0 & 0.01 & - \\
        4.77  & 0.28 & 3.1 & 0.04 & - \\
        9.55  & 0.27 & 1.2 & 0.00 & 0.16 \\
        43.91 & 0.27 & 1.8 & 0.06 & 0.21 \\
        9.55  & 0.28 & 1.8 & 0.01 & 0.57 \\
        \hline
    \end{tabular}
    \caption{Fragmentation events for Case 7.}    
    \label{tab:case7}
\end{table}

\begin{table}
    \centering
    \begin{tabular}{|c|c|c|c|c|}
        \hline
        Mass & $P_\text{release}$ & $C_\text{fr}$ & $\alpha$ & $\sigma_0$ \\
        $\text{[ton]}$ & [MPa] & & & [MPa] \\
        \hline
        \hline
        9.55  & 0.10 & 2.0 & 0.01 & 0.24 \\
        9.55  & 0.12 & 2.1 & 0.00 & 0.46 \\
        7.64  & 0.52 & 2.0 & 0.00 & - \\
        11.46 & 0.80 & 2.0 & 0.00 & - \\
        9.55  & 1.09 & 2.0 & 0.03 & - \\
        7.64  & 1.45 & 1.9 & 0.00 & - \\
        11.46 & 0.90 & 1.1 & 0.00 & 1.35 \\
        42.00 & 0.90 & 2.0 & 0.06 & 1.40 \\
        7.64  & 0.90 & 2.0 & 0.01 & 3.90 \\
        \hline
    \end{tabular}
    \caption{Fragmentation events for Case 8.}    
    \label{tab:case8}
\end{table}

\begin{table}
    \centering
    \begin{tabular}{|c|c|c|c|c|}
        \hline
        Mass & $P_\text{release}$ & $C_\text{fr}$ & $\alpha$ & $\sigma_0$ \\
        $\text{[ton]}$ & [MPa] & & & [MPa] \\
        \hline
        \hline
        1.91  & 1.30 & 2.0 & 0.00 & 2.80 \\
        5.73  & 2.70 & 2.5 & 0.00 & 3.70 \\
        3.82  & 6.00 & 3.0 & 0.02 & - \\
        5.73  & 5.90 & 1.4 & 0.01 & - \\
        7.64  & 6.50 & 2.4 & 0.01 & 11.00 \\
        45.82 & 6.50 & 2.4 & 0.05 & 11.20 \\
        17.18 & 6.90 & 2.1 & 0.00 & 23.40 \\
        \hline
    \end{tabular}
    \caption{Fragmentation events for Case 9.}
    \label{tab:case9}
\end{table}


\bsp	
\label{lastpage}
\end{document}